\documentclass[a4paper,11pt]{JHEP3}
\usepackage{graphicx}
\usepackage{epsfig}
\usepackage{amsmath, amssymb}
\usepackage{dcolumn}
\usepackage{bm}
\usepackage{amsfonts}

\usepackage{subfigure}
\usepackage{amssymb}

\newcommand{\dd}{\partial}

\newcommand{\be}{\begin{equation}}
\newcommand{\ee}{\end{equation}}
\newcommand{\ben}{\begin{equation*}}
\newcommand{\een}{\end{equation*}}
\newcommand{\bea}{\begin{eqnarray}}
\newcommand{\eea}{\end{eqnarray}}
\newcommand{\bean}{\begin{eqnarray*}}
\newcommand{\eean}{\end{eqnarray*}}
\newcommand{\brr}{\begin{array}}
\newcommand{\err}{\end{array}}
\newcommand{\bc}{\begin{center}}
\newcommand{\ec}{\end{center}}

\newcommand{\lsim}{\,\raisebox{-0.6ex}{$\buildrel < \over \sim$}\,}
\newcommand{\gsim}{\,\raisebox{-0.6ex}{$\buildrel > \over \sim$}\,}

\newcommand{\bB}{{\mathbf B}}

\newcommand{\bn}{{\mathbf n}}

\newcommand{\al}{\alpha}

\newcommand{\de}{\delta}
\newcommand{\De}{\Delta}

\newcommand{\ga}{\gamma}

\newcommand{\Om}{\Omega}
\newcommand{\om}{\omega}

\newcommand{\lp}{\left}
\newcommand{\rp}{\right}
\newcommand{\ape}{a_\perp}
\renewcommand{\apa}{a_\parallel}

\title{A large scale coherent magnetic field:\\  interactions with free 
streaming particles and limits from the CMB}
\author{Julian Adamek
\\ Institut f\"ur theoretische Physik und Astrophysik, Universit\"at W\"urzburg, Am Hubland, 97074 W\"urzburg, Germany\\
 \email{jadamek@physik.uni-wuerzburg.de} 
}
  \author{Ruth Durrer  
 \\ Institut de Physique Th\'eorique, Universit\'e de Gen\`eve, 24 quai 
E. Ansermet, 1211 Gen\`eve 4, Switzerland\\ 
  \email{ruth.durrer@unige.ch} 
  }
 \author{Elisa Fenu
 \\ Institut de Physique Th\'eorique, Universit\'e de Gen\`eve, 24 quai 
E. Ansermet, 1211 Gen\`eve 4, Switzerland\\ 
  \email{elisa.fenu@unige.ch} 
  }
 \author{Marc Vonlanthen  
 \\ Institut de Physique Th\'eorique, Universit\'e de Gen\`eve, 24 quai 
E. Ansermet, 1211 Gen\`eve 4, Switzerland\\ 
  \email{marc.vonlanthen@unige.ch}
}
\received{\today}


\abstract{We study a 
homogeneous and nearly-isotropic Universe permeated by a homogeneous magnetic field.
Together with an isotropic fluid, the homogeneous magnetic 
field, which is the primary source of anisotropy, leads to a plane-symmetric Bianchi~I model of the Universe.
However, when free-streaming relativistic particles are present, they generate an anisotropic pressure which 
counteracts the one from the magnetic field such that the Universe becomes 
isotropized. We show that due to this effect, the CMB  temperature anisotropy 
from a homogeneous magnetic field is significantly suppressed if the the neutrino masses
are smaller than 0.3 eV.
}

\maketitle

\begin{document}


\section{Introduction}\label{S:intro}


On very large scales, the observed Universe is well approximated by a 
homogeneous and isotropic Friedmann solution of Einstein's 
equations. This is best verified by the isotropy of the Cosmic Microwave Background
(CMB). The small fluctuations observed in the CMB temperature are fully accounted for
by the standard model of structure formation from small initial fluctuations
which are generated during an inflationary phase.  Nevertheless, these small 
fluctuations are often used to limit other processes or components which may 
be present in the early Universe, like e.g.\ a primordial magnetic field. 

The generation of the magnetic fields observed in galaxies and clusters~\cite{obs} is
still unclear. It has been shown that phase transitions in the early Universe, 
even if they do generate magnetic fields, have not enough power on large 
scale to explain the observed large scale coherent fields~\cite{Btrafo}.
These findings suggest that primordial magnetic fields must be
correlated over very large scales.

In this paper, we discuss limits on fields which are coherent over a Hubble 
scale and which we can therefore treat as a homogeneous magnetic field 
permeating the entire Universe. We want to derive limits on a homogeneous field
from CMB anisotropies.
This question has been addressed in the past~\cite{Barrow1} and limits on the 
order of $B\lsim 2\times 10^{-9}$ Gauss have been derived from the CMB 
anisotropies~\cite{Barrow2}. A similar limit can also be obtained from
Faraday rotation~\cite{fara,wmap}.

We show that the limits from the CMB temperature anisotropy
actually are invalid if free streaming neutrinos with masses 
$m_\nu<T_{\rm dec}$ are present, where $T_{\rm dec}$ denotes the 
photon temperature at decoupling. This is the case if the neutrino 
masses are not degenerate, i.e.\ the largest measured mass splitting is of
the order of the largest mass, hence $m_\nu \lsim 0.04$\rm eV.
The same effect can be obtained from any other massless free 
streaming particle species, like e.g.\ gravitons, if they contribute 
sufficiently to the background energy density. This is due to the following
mechanism which we derive in detal in this paper:
In an anisotropic Bianchi-I model, free streaming relativistic particles 
develop an anisotropic stress. If the geometric anisotropy is due to a
magnetic field, which scales exactly like the anisotropic stress of the massless
particles, this anisotropic stress cancels the one from the magnetic field and
the Universe is isotropized. Hence the quadrupole anisotropy of the CMB 
due to the magnetic field is erased. This `compensation' of the magnetic field
anisotropic stress by free-streaming neutrinos has also been seen in the study 
of the effects of stochastic magnetic fields on the 
CMB~\cite{Finelli,ShawLewis,CC,Kerstin}
for the large scale modes. In our simple analysis the mechanism behind
it finally becomes clear.
  
The limits from Faraday rotation are not affected by our arguments.

In the next section we derive the CMB anisotropies in a Bianchi~I Universe.
In Section~\ref{S:nus} we show that relativistic free streaming neutrinos
in a Bianchi~I model develop anisotropic stresses and that these back-react to 
remove the anisotropy of the Universe if the latter is due to a massless mode. 
In Section~\ref{S:GWs} we discuss isotropization due
to other massless free streaming particles, with special attention to a
gravitational wave background. In Section~\ref{S:con} we conclude.

\section{Effects on the CMB from a constant magnetic field in an ideal fluid
Universe}\label{S:onCMB}

We consider a homogeneous magnetic field in $z-$direction, $\bB=B\bm{e}_z$
in a Universe filled otherwise with an isotropic fluid consisting, e.g.\ of matter and 
radiation. The metric of such a Universe is of Bianchi type I,
\bea 
ds^2 &=&-dt^2+\ape^2(t)(dx^2+dy^2)+\apa^2(t)\,dz^2 \,,	
\eea
where $t$ is cosmic time. 
The Einstein equations in cosmic time read
\bea
2 \frac{\dot{a}_\parallel}{\apa} \frac{\dot{a}_\perp}{\ape} + \lp(\frac{\dot{a}_\perp}{\ape}\rp)^2 &=& 8 \pi G \rho~,\label{EE1}\\
\frac{\ddot{a}_\parallel}{\apa} + \frac{\ddot{a}_\perp}{\ape} + \frac{\dot{a}_\parallel}{\apa} \frac{\dot{a}_\perp}{\ape} &=& -8 \pi G P_\perp~,\label{EE3}\\
2 \frac{\ddot{a}_\perp}{\ape} + \lp(\frac{\dot{a}_\perp}{\ape}\rp)^2 &=& -8 \pi G P_\parallel~.\label{EE2}
\eea

The dot denotes the derivative with respect to $t$. We have introduced the total energy density
$\rho = \rho_B + \rho_m + \rho_\gamma + \rho_\nu + \rho_\Lambda$, where $\rho_B = B^2/8 \pi$
is the energy density in the magnetic field, and $\rho_m, \rho_\gamma, \rho_\nu, \rho_\Lambda$
are as usual the energy densities of matter (assumed to be baryons and cold dark matter), photons,
neutrinos, and dark energy (assumed to be a cosmological constant), respectively.

All the above constituents of the Universe, except matter (which is assumed to be pressureless) also
contribute to the pressure components $P_\parallel, P_\perp$. The contribution from the magnetic field
is intrinsically anisotropic and given by
\be \label{e:PB}
 P_{B, \perp} = - P_{B, \parallel} = \rho_B\,,
 \ee
as can be read off from the corresponding stress-energy tensor. 
Note that the magnetic field $B$ decays as $\ape^{-2}$, so that 
$\rho_B$ scales as $\ape^{-4}$.

For later reference we define an `average' scale factor 
\be
a \equiv \ape^{2/3} \apa^{1/3}\,,
\ee
which is chosen such that it correctly describes the volume expansion.

Let us also introduce the expansion rates $H_\perp = \dot{a}_\perp / \ape$ and 
$H_\parallel = \dot{a}_\parallel / \apa$.
The anisotropic stress of the homogeneous magnetic field sources anisotropic
expansion, which can be expressed as the difference of the expansion 
rates, $\Delta H = H_\perp - H_\parallel$. We combine 
eqs.~(\ref{EE2}) and (\ref{EE3})
to obtain an evolution equation for $\Delta H$,
\be
\dot{\Delta H} + \lp(2 H_\perp + H_\parallel\rp) \Delta H = 8 \pi G \lp(P_\perp - P_\parallel\rp)\,.\label{e:deltaH}
\ee

This pressure difference is actually simply the anisotropic stress. More 
precisely,
\bea
{\Pi_i}^j &\equiv& {T_i}^j-P{\de_i}^j \,, \qquad P= T_i^i/3 = (2P_\perp+
  P_\parallel)/3\,, \nonumber \\
{\Pi_1}^1 &=& {\Pi_2}^2 = 
P_\perp -P = \frac{1}{3}\left(P_\perp-P_\parallel\right) \,, \quad
{\Pi_3}^3 = P_\parallel -P = -\frac{2}{3}\left(P_\perp-P_\parallel\right)\,.
\eea
At very high temperatures, both photons and neutrinos are tightly 
coupled to baryons. Their pressure is isotropic and thus their contribution 
to the right-hand-side of (\ref{e:deltaH}) vanishes. The collision term in
Boltzmann's equation tends to isotropize their momentum-space distribution. 
Under these conditions the only source of anisotropic stress is the magnetic field. 
The above equation can then easily be solved to leading order
in $\Delta H$, as will be carried out in section~\ref{S:nus}.

However, as soon as the neutrinos decouple and start to free-stream, their 
momentum-space distribution will be affected by the anisotropic expansion 
caused by the magnetic field and thus they will develop anisotropic stress. As we
will show, the neutrino anisotropic stress counteracts the one from the magnetic 
field. This behavior will be maintained until the neutrinos become 
non-relativistic, then their pressure decays. For the temperature
anisotropy in the CMB it is relevant whether this happens before or after 
photon decoupling. This depends, of course,
on the neutrino masses.

We introduce the energy density parameters 
$$\Omega_x(t) \equiv \frac{8\pi G \rho_x(t)}{3H^2(t)} = \frac{ \rho_x(t)}{ \rho_c(t)} \;,$$ 
corresponding respectively to the magnetic field, matter and radiation etc., such  that
e.g. $ \Om_B=B^2/8 \pi\rho_{c}$,
 $\Om_m=\rho_m/\rho_{c}$ and 
$\Om_\gamma=\rho_\gamma/\rho_{c}$.   
Here we define the `average' Hubble parameter by
\be
 H^2\equiv \frac{1}{3}\lp[ \lp(\frac{\dot{a}_\perp}{\ape}\rp)^2+
2\frac{\dot{a}_\perp\dot{a}_\parallel}{\ape\apa}\rp] \;. \ee
With this,  eq.~(\ref{EE1}), implies  
\be\label{e:const}
\Omega_{\rm T} \equiv \Om_B + \Om_\gamma + \Om_\nu + \Om_m +
\Om_\Lambda=1 \quad \mbox{ at all times.}
\ee
As an alternative, one could have defined the `average' Hubble parameter as
$$H_a \equiv \frac{1}{3}\lp[
2\frac{\dot{a}_\perp}{\ape}+\frac{\dot{a}_\parallel}{\apa}\rp]\;. $$
It can easily be verified that the difference between these definitions is
of the order of the small quantity
$\De H = H_\perp -H_\parallel$. More precisely, 
\bea 
H^2   &=& H_a^2\left[1 - \frac{2}{3}\frac{\De H}{H_a} - 
            \frac{1}{3}\left(\frac{\De H}{H_a}\right)^2\right] \,.
\eea
We shall mainly use the definition which yields the constraint~(\ref{e:const}).

The scaling of the energy densities corresponding to every species 
follows from the stress energy conservation of every single fluid
\be 
   \rho_\ga=\rho_\gamma^0 \lp(\frac{a_0}{a} \rp)^4\,, \quad  
  \rho_m=\rho_m^0 \lp(\frac{a_0}{a} \rp)^3\,, 
   	\quad \rho_B=\rho_B^0 \lp[\frac{\ape(t_0)}{\ape} \rp]^4\,.
\ee
To obtain the above behavior for radiation, it is important to impose that
the fluid is ideal, i.e.\ that pressure is isotropic. This is the case if 
there are sufficiently many collisions, but does not hold for free streaming
particles as we shall see in the next section.

At a fixed initial time one may set $\ape=\apa$ as initial condition. 
Motivated by observations, we assume that the scale factor difference always remains small,
\be\label{e:deltadef}
 \frac{\ape-\apa}{a} \equiv \de \ll 1 \;.
\ee 
To first order in $\De H \ll H$, as long as the magnetic field is the only
anisotropic component, eq.~(\ref{e:deltaH}) becomes (see also~\cite{GHR})
\begin{equation}
\label{GHReq}
 \dot{\Delta H} + 3 H \Delta H = 8 \pi G \left(P_\perp - P_\parallel\right)
=6H^2\Om_B \,.
\end{equation}

In the following we consider both $\Om_B$ and $\De H$ as small quantities
and want to calculate effects to first order in them. To first order, 
$\rho_B \propto a^{-4} \propto \rho_\ga$. We can therefore 
introduce the ratio
\be  \label{e:r}
 r = \frac{\rho_B}{\rho_\ga} = \frac{\Om_B}{\Om_\ga}\,,
 \ee
which (to first order) is constant.

\begin{figure}[ht]
\centering  
\includegraphics[width=12cm]{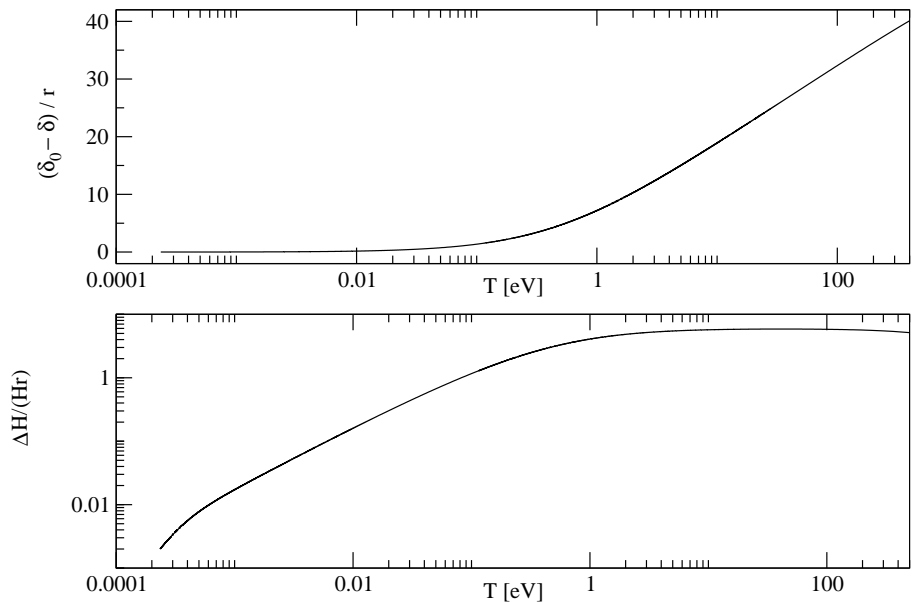}
\caption{\label{f:scalef}
 Temperature evolution of the scale factor difference $\de_0-\de$ and $\De H/H$
 in units of $r=\Om_B/\Om_\ga$ when no free-streaming particle compensates the anisotropy produced by the magnetic field anisotropic stress.
 Here $\de_0$ denotes the scale factor difference $\de$ today.
The evolution of the `average' scale factor $a$ is the one of a $\Lambda$CDM Universe. 
As it is shown in section 3, $\De H/H$ is constant during the radiation dominated 
era and $\de$ is growing. During the matter dominated era  $\De H/H$ is decaying,
$\De H/H\propto 1/a \propto T $, and $\de$ asymptotes to a constant.}
\end{figure}

In fig.~\ref{f:scalef} we plot the scale factor difference $\de_0-\de$ and $\De H/H$
as functions of the temperature in a first stage where neutrinos, photons and baryons are all tightly coupled and the magnetic field is the only source of anisotropy. 

\subsection{Lightlike geodesics in Bianchi~I}
Let us now determine the CMB anisotropies in a
Bianchi~I Universe. We are not interested in the usual anisotropies from 
primordial perturbations, which we disregard in our treatment, but we
concentrate on the effect of the global anisotropy, which to leading order
will result in a temperature quadrupole. 

 We choose the tetrad basis $e_0=\dd_t$, $e_i=\ape^{-1}\dd_i$ for $i=1,2$ and
$e_3=\apa^{-1}\dd_3$. The dual basis of 1-forms is given by
$\theta^0=dt$, $\theta^i=\ape dx^i$, for $i=1,2$ and $\theta^3=\apa dx^3$.
The first structure equation,
$$ d\theta^a +{\om^a}_b\wedge\theta^b = 0  \,,$$
yields
\bea
\label{omegas1}
{\om^i}_0 &=& \frac{\dot\ape}{\ape}\theta^0\;, \quad i=1,2 \,,\qquad {\rm and} \quad
{\om^3}_0 = \frac{\dot\apa}{\apa}\theta^0\,.
\eea
The other non-vanishing connection 1-forms are determined by anti-symmetry, 
$\om_{ab}=-\om_{ba}$. 
After photon decoupling, the photon 4-momentum ${\bm p}=p^a\bm{e}_a$ satisfies 
the geodesic equation
\be
  \frac{d p^a}{d \lambda} + \omega_{\phantom{1} c}^{a}(\bm e_b) \,p^b\, p^c
 =0  \,.
\ee
Considering the constraint relation for massless particles $p_a p^a=0$ and 
setting $\alpha T_0\equiv p^0 =p = \sqrt{\sum_{i=1}^3 (p^i)^2 }$, where 
$T_0$ is a constant with the dimension of energy (or temperature) that 
multiplies all the components $p^a$, the above equation is solved by
\be
\label{p}
 \lp(p^a\rp) = T_0\lp(\alpha, \frac{n^1}{\ape}, \frac{n^2}{\ape},
       \frac{n^3}{\apa}\rp) \,,
\ee
where $n$ is a unit vector in the direction of the particle momentum and $\alpha$ is 
determined by the condition $p_ap^a=0$.
$$ n^1 =\sin\theta\sin\phi\;, \quad n^2=\sin\theta\cos\phi
 	\quad \mbox{ and }\quad n^3=\cos\theta \;.
$$
The temperature of photons in such an anisotropic Universe for a comoving
observer, $u=\dd_t$, is then given by
\be
\label{T}
T(t,\theta) = \eta_{ab}u^ap^b = p^0 =T_0\al = T_0\sqrt{\frac{\sin^2\theta}{\ape^2} +
   \frac{\cos^2\theta}{\apa^2}} \simeq \frac{T_0}{a}\lp[1 +\de\cos^2\theta +
	{\cal O}(\de^2)\rp]\,.
\ee  
We set
$$ \bar{T} = \frac{1}{4\pi}\int T(t,\theta) \sin\theta d\theta d\phi
  = \frac{T_0}{a}\lp[ 1+ \frac{1}{3}\de + {\cal O}(\de^2)\rp]
$$
to be the temperature averaged over directions.
Note that for $\de =0$ and $a_0=1$, $T_0$ is simply the CMB 
temperature at time $t_0$. For the temperature fluctuations to 
first order in $\de$  we obtain
\be
\frac{\De T}{T} \equiv \frac{T(t,\theta)-\bar T}{\bar T} = 
  \frac{1}{3}\de(3\cos^2\theta-1) + {\cal O}(\de^2)
=\de\frac{2}{3}\sqrt{\frac{4\pi}{5}}Y_{20}(\bn) + {\cal O}(\de^2) \;.
\ee
Hence, to lowest order in $\de$ a homogeneous magnetic field generates
a quadrupole which is given by
\be\label{e:quad}
C_2 = \frac{1}{5}\sum_{m=-2}^2|a_{2m}|^2 = \frac{1}{5}|a_{20}|^2 = 
\frac{16\pi}{225}\de^2 \simeq 0.22\times\de^2\;.
\ee

Of course, in principle one can set $\de(t_1)=0$ at any given moment $t_1$
which then leads to $\frac{\De T}{T}(t_1) =0$. However, for the CMB we 
know that photons start free-streaming only at $t_{\rm dec}$ when they decouple from electrons. 
Before that, scattering isotropizes the photon 
distribution and no quadrupole can develop\footnote{This is not strictly true and 
neglects the slight anisotropy of non-relativistic Thomson scattering.}.
In other words, we have to make sure that the anisotropy-induced quadrupole
is fixed to zero at decoupling and only appears as a result of differential
expansion between last scattering and today.
This can be taken into account by simply choosing the 
initial condition  $\de(t_{\rm dec})=0$. Without this initial condition we 
have to replace $\de(t)$ by $\de(t)-\de(t_{\rm dec})$ in eq.~(\ref{e:quad})
\footnote{More generally, one can say that $\de$ itself is not a quantity with
a physical meaning as long as no reference value is specified. In physical terms,
only the difference of $\de$ between two instants of time can be a relevant quantity.}.
The general result for the CMB quadrupole today is therefore
\be\label{e:quad2}
C_2 = \frac{16\pi}{225}\left[\de(t_0) -\de(t_{\rm dec})\right]^2 \;.
\ee

\subsection{The Liouville equation}

At this stage it is straightforward to check that the exact expression found 
above for the temperature, eq.~(\ref{T}), satisfies the Liouville equation 
for photons (see, e.g. ~\cite{Durrer:1993db})
\be
p^{a} \bm e_a (f_\gamma) - \omega^{i}_{ \phantom{1}b}({ \bm p})  p^b 
\frac{\partial f_\gamma}{\partial p^i} =0\,,
\ee
when we make the following Ansatz for the distribution function of massless 
bosonic particles in our Bianchi~I Universe
\bea
 && p_\bot \equiv \sqrt{(p^1)^2+(p^2)^2}\,, \quad p_\parallel = p^3, \quad 
 p = \sqrt{ p_\bot^2 + p_\parallel^2} = p^0\,,  \\
 && f_\gamma(t, T) =\frac{N_\gamma}{(2\pi)^3} \frac{1}{{\rm e}^{p/T}-1}, \qquad  
 T=T(t,\theta)\,. \label{fbose}
\eea
Indeed, using eqs.~(\ref{omegas1}), we find the following differential 
equation for the temperature $T$
\be
\frac{\partial f_\gamma}{\partial T}\frac{\partial T}{\partial t} - 
\frac{\dot \ape}{\ape}\frac{\partial f_\gamma}{p_\bot} p_\bot- 
\frac{\dot \apa}{\apa} \frac{\partial f_\gamma}
		{\partial p_\parallel}p_\parallel=0 \,.
\ee
With (\ref{fbose}) this can be written as
\be
	\frac{\dot T}{T} + \frac{\dot \ape}{\ape} \lp(\frac{p_\bot}{p}\rp)^2 + 
	\frac{\dot \apa}{\apa} \lp(\frac{p_\parallel}{p}\rp)^2=0 \,.
\ee
The time behavior of the different components of the photon momentum are given 
by eq.~(\ref{p}) and one immediately sees that expression~(\ref{T}) for the 
temperature solves the above differential equation.

Moreover, defining the time dependent unit vectors $\hat p^i \equiv p^i/p$ and the shear tensor 
$$\sigma_{ab} \equiv  \vartheta_{ab}-\frac{1}{3} \vartheta_{c}^{c} h_{ab} \ , ~~ \mbox{  where  }
\vartheta_{ab} \equiv \frac{1}{2}\left( \nabla_a u_b + \nabla_b u_a\right)  ~ \mbox{  and } ~
h_{ab}\equiv \eta_{ab}+u_a u_b\, , $$ 
one can rewrite  the above Liouville equation as
\be
	(\tilde{p})^{\dot{}} =-\tilde{p}\sigma_{ij} \hat p^i \hat p^j\,,
\ee
where $\tilde{p}$ denotes the redshift-corrected photon energy 
defined as $\tilde{p}\equiv a p$. This last expression agrees with 
the corresponding equation given in~\cite{Pontzen:2007ii}.

Using the expression for the distribution function of massless fermions, 
we can also compute the pressure of neutrinos once they start free-streaming. 
Indeed, given the fact that neutrinos can be considered massless before they 
become non-relativistic, their geodesic equation has the same solution as the 
one for photons found above, therefore we immediately obtain 
the time behavior of their temperature  in an anisotropic Bianchi I background.
Taking also into account the fact that neutrinos are fermions, their 
distribution function reads
\be
f_\nu(t, T) =\frac{N_\nu}{(2\pi)^3} \frac{1}{{\rm e}^{p/T}+1}, \quad \mbox{with}
 \quad  T(t,\theta)=\frac{T_\nu}{a}\lp[1 +\de\cos^2\theta +
  {\cal O}(\de^2)\rp]\,.
\ee
Note that the parameter $T$ appearing in the neutrino distribution function 
in not a temperature in the thermodynamical sense as the neutrinos are not 
in thermal equilibrium. It is simply a parameter in the distribution function and its
time evolution has been determined by requiring the neutrinos to move along 
geodesics i.e. to free-stream.

This distribution function remains valid also in the case where neutrinos are
massive, i.e. $T_\nu < m_\nu$. The only difference is that  the relation $p^0=p$
changes to $p^0 = \sqrt{p^2 + (m_\nu a)^2}$ which of course affects the 
momentum integrals for the neutrino energy density and pressure.
 
  The energy $T_\nu/a_0$
is the present neutrino `temperature' in the absence of a homogeneous
magnetic field ($\de =0$). The energy density $\rho_\nu$ and the pressure $P_{\nu,i}$ in direction $i$ 
with respect to our orthonormal basis are
\bea 
\rho_{\nu} &=&  N_\nu\int d^3p \,f_\nu(t, T)p^0 \\  \label{e:pressure}
	P_{\nu,i} &=& N_\nu\int d^3p \,f_\nu(t, T)\, \frac{p_i^2}{p^0}\,.
\eea

Calculating the integral (\ref{e:pressure}) for relativistic neutrinos to first order 
in $\delta$ in the directions perpendicular and parallel to the magnetic field 
direction, one finds for the neutrino anisotropic stress in the ultra-relativistic limit
\be
\label{nuPress}
  P_{\nu, \perp} - P_{\nu, \parallel} \simeq -\frac{8}{15}\rho_\nu \lp(\delta - 
  \delta_\ast\rp)\,,
\ee
where $\delta_\ast$ is the value of $\delta$ at neutrino decoupling and can 
be fixed to zero for convenience.

\begin{figure}[ht]
\centering   
\vspace{1cm}  ~  \\
\includegraphics[width=12cm]{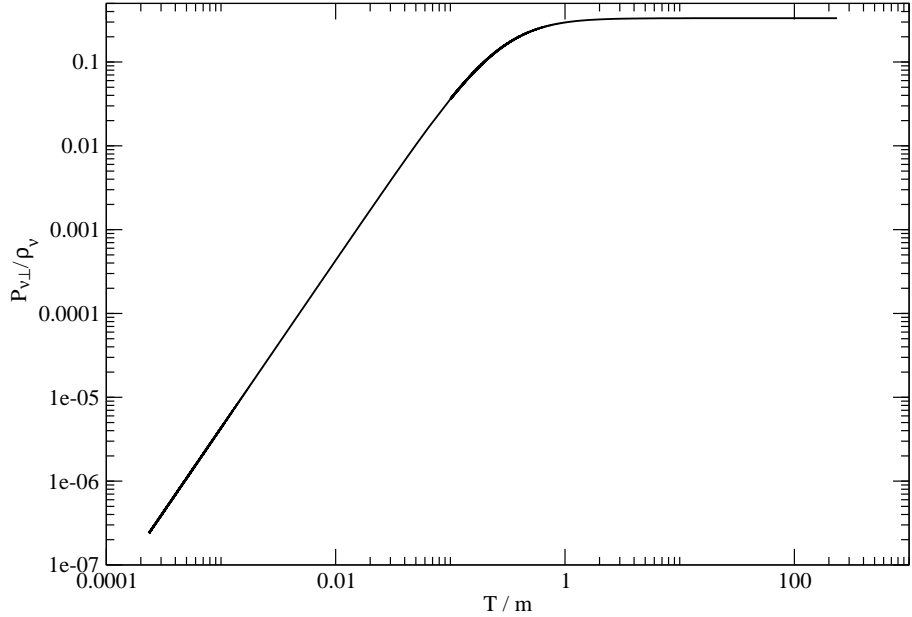}
\caption{\label{f:pressure}Temperature evolution of the neutrino pressure
$P_{\nu, \perp}$ normalized to the neutrino energy density $\rho_\nu$. The temperature is given in units
of the neutrino mass. Note that the break in the power law is not at $T=m$, but at somewhat lower temperature.
This is due to the highly relativistic Fermi-Dirac distribution of the neutrinos, see also the discussion of
fig.~5 in sec.~3.3.}
\end{figure}

The temperature dependence of the neutrino pressure is shown in fig.~\ref{f:pressure}. To leading
order, this also gives the temperature dependence of the neutrino anisotropic stress. From the plot
it is clear how the pressure scales as $a^{-4}$ as long as the neutrinos are ultra-relativistic.
Once they have become effectively non-relativistic, their pressure decays more rapidly, as $a^{-5}$.
The break in the power law is not precisely at $T = m_\nu$, but at a somewhat lower temperature.
Because the neutrinos still have the highly relativistic Fermi-Dirac distribution from the time of their
thermal freeze-out, it takes some additional redshift until they behave effectively non-relativisic.
This will have some effect on the estimates for the residual CMB quadrupole, as we shall see 
in sec.~\ref{S:nus}, in particular the discussion of fig.~\ref{f:quadru}.

\section{Neutrino free-streaming and isotropization}\label{S:nus}

\subsection{Massless free-streaming neutrinos}

We now calculate the effect of free-streaming neutrinos perturbatively, i.e.\ to first order
in $\de$, $\De H/H$ and $\Om_B$. We linearize 
eq.~(\ref{e:deltaH}), taking into account the contribution of a free-streaming 
relativistic component to the right-hand side.
We have shown that this contribution, to leading order in $\delta$, is given 
by eq.~(\ref{nuPress}). Furthermore, up to $\mathcal{O}(\delta^2)$ corrections, $\delta$ 
is just the integral of $\Delta H$,
\be
\int_{t_\ast}^t\!\Delta H(t') d t' = \ln \frac{\ape\lp(t\rp)}{\apa\lp(t\rp)} - 
\ln \frac{\ape\lp(t_\ast\rp)}{\apa\lp(t_\ast\rp)} \simeq \delta - \delta_\ast~,
\ee
so that to first order we can identify $\Delta H \simeq \dot{\delta}$.

Inserting this back into eq.~(\ref{e:deltaH}) we find, to linear order in 
$\delta$,
\be \label{e:lineardelta}
\ddot{\delta} + 3 H \dot{\delta} +\frac{8}{5} H^2 \Omega_\nu \lp(\delta - 
\delta_\ast\rp) = 6 H^2 \Omega_B~. 
\ee
Note that, because we are working at linear order, it is not important with 
respect to which scale factor $H, \Omega_\nu$ and $\Omega_B$ are 
defined in (\ref{e:lineardelta}).
We will now give analytic solutions to this equation for different regimes in the 
evolution of the Universe.

Let us begin at very high temperature where the neutrinos are still strongly coupled to baryons.
In this case they do not contribute to eq.~(\ref{e:lineardelta}) since their pressure is isotropic $(P_{\nu, \perp} - P_{\nu, \parallel}\sim0)$ given the high rate of collisions. Furthermore, since we are in the {\bf radiation
dominated era} $(a\propto t^{1/2})$, we have $H = 1/2t$, and $\Omega_B$ is constant. The solution to eq.~(\ref{e:lineardelta}) in
this case is
\be
\dot{\delta}(t) = \Delta H (t) = \frac{3 \Omega_B}{t} + \frac{C}{t^{3/2}}~.
\ee
The dimensionless quantity $\Delta H / H$ hence asymptotes to a constant, 
since the homogeneous piece decays like $a^{-1}$:
\be
\frac{\Delta H}{H} \rightarrow 6 \Omega_B~. \label{e:RDlimit}
\ee
$\Delta H$ soon becomes insensitive to the initial conditions and only depends on $\Omega_B$.
This also shows that in the absence of an anisotropic source ($\Om_B=0$), 
the expanding Universe isotropizes. Integrating this equation and remembering that 
$\Om_B =$ constant to first order in a radiation dominated Universe, we obtain
\be
\label{e:RDlimitDe}
\de(t)-\de(t') = 3\Om_B\ln(t/t') \,.
\ee

\begin{figure}
\centering   
\includegraphics[width=12cm]{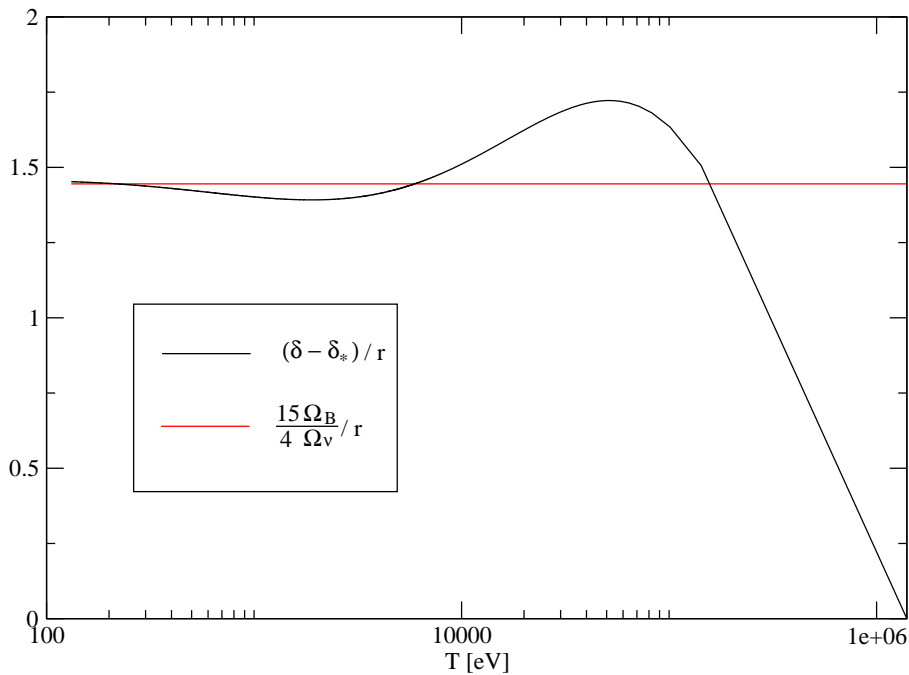}
\caption{
Temperature evolution of $\delta-\delta_\ast$ from neutrino decoupling to $T=100 \mathrm{eV}$. After
decoupling, $\delta-\delta_\ast$ begins to oscillate with a decreasing amplitude around the constant 
$\frac{15}{4} \frac{\Omega_B}{\Omega_\nu}$, as predicted by the analytic solution~(\protect\ref{e:RDsolution}). 
This qualitative behavior is independent of the initial conditions.
\label{f:del_nudec} }
\end{figure}

As the Universe reaches a temperature of roughly $1.4$ MeV, the neutrinos
decouple and begin to free-stream, giving rise to the corresponding term in 
eq.~(\ref{e:lineardelta}). In the radiation dominated era, $\Omega_\nu$
remains constant as long as neutrinos are ultra-relativistic\footnote{Actually, 
$\Omega_\nu$ changes slightly when electron-positron annihilation takes place, 
a process which heats up the photons but not the neutrinos. This happens at a
temperature close to the electron mass. After that, $\Omega_\nu /\Om_\gamma$ 
 remains constant until the neutrinos become non-relativistic.}. This is 
certainly true for temperatures well above a few eV. In this regime, the 
general solution of eq.~(\ref{e:lineardelta}) is given by
\be
\label{e:RDsolution}
\delta(t) - \delta_\ast = \frac{15}{4} \frac{\Omega_B}{\Omega_\nu} + t^{-1/4} 
\lp(C_+ t^{i \sqrt{2\Omega_\nu/5 - 1/16}} + C_- t^{-i \sqrt{2\Omega_\nu/5 - 1/16}}\rp)~.
\ee
 For $\Omega_\nu > 5/32$, the homogeneous part is oscillating with a damping 
 envelope $\propto t^{-1/4} \propto a^{-1/2}$.
This means that $\Delta H = \dot{\delta}$ will decay within a few Hubble times,
which is a matter of seconds at the temperatures we are talking about. After 
that, $\delta - \delta_\ast$ will remain constant at the value of $\lp(15 / 
4\rp) \Omega_B / \Omega_\nu$ until the neutrinos become non-relativistic. 
Then their pressure drops dramatically and so does their anisotropic stress. 
Until this time, the Universe expands isotropically, because the anisotropic
stress of the magnetic field is precisely cancelled by the one of the 
neutrinos. Remember that a constant $\de$ can always be absorbed in a re-scaling 
of the coordinates and has no physical effect. Fig.~\ref{f:del_nudec} shows 
the temperature evolution of $\delta - \delta_\ast$ in the radiation dominated 
era from neutrino decoupling until $T=100 \mathrm{eV}$.

This mechanism rests on two important facts.  Firstly, as long as neutrinos are 
ultra-relativistic, they redshift in the same way as the magnetic field, 
meaning that $\Omega_B / \Omega_\nu$ is constant. Once the anisotropic
stress of the neutrinos has adjusted to the magnetic field, their sum remains zero 
independent of the expansion of the Universe which is now in a Friedmann phase. 
Secondly, the efficiency of the effect hinges on the absolute value of $\Omega_\nu$. 
In the radiation dominated era (after positron annihilation), we have 
$\Omega_\nu \simeq 0.4$ so that $\Omega_\nu  > 5 /32$,
and hence the system behaves as an underdamped oscillator with a damping 
envelope $\propto t^{-1/4}$. Had the density parameter of the free-streaming 
particles been less than $5/32$, the behavior would be that of an overdamped 
oscillator. As it is evident from eq.~(\ref{e:RDsolution}), for 
$\Omega_\nu \ll 5/32$ there would be a mode which decays extremely slowly,
roughly as $t^{-4\Omega_\nu / 5}$. This is why a strongly subdominant 
free-streaming component cannot damp the anisotropy efficiently. As we shall 
discuss in  section~\ref{S:GWs}, a primordial gravitational wave background 
could play the role of such a  free-streaming component if 
$\Omega_{GW} \gtrsim 5/32$.

\subsection{Massive neutrinos}

The neutrinos become non-relativistic roughly at the time when their 
temperature drops below their mass scale. Current bounds on the neutrino 
mass \cite{numass} are such that the highest-mass eigenstate is somewhere 
between $\sim 1$ eV and $\sim 0.04$ eV. Since the neutrino mass splitting is 
much below $1$ eV, an eigenstate close to the upper bound would mean that the 
neutrinos are almost degenerate and hence become non-relativistic all at the 
same time. If this happens \textit{before} photon decoupling, i.e., if
$m_\nu>0.3$ eV, the isotropization effect
will not be present and the CMB will be affected by the anisotropic expansion sourced by the magnetic field. However,
if the neutrinos remain ultra-relativistic until long \textit{after} photon decoupling, 
the CMB quadrupole due to anisotropic expansion will be reduced because the 
neutrinos maintain expansion isotropic until they become
non-relativistic.

In order to quantify this statement, we repeat the above calculations 
for the {\bf matter dominated era}. For our purposes,
this is a reasonable approximation for the time between photon 
decoupling and today. At decoupling, radiation
is already subdominant, and  on the other hand vacuum 
energy only begins to dominate at redshift $z\sim 0.5$. We 
therefore expect that both give small corrections only.

For completeness, we also give the solution of eq.~(\ref{e:lineardelta}) 
in a matter dominated Universe for the case where we ignore any contributions
from free-streaming particles (neutrinos and, after decoupling, 
also photons). During matter domination we have $H = 2/3t$ and 
$\Omega_B \propto a^{-1} \propto t^{-2/3}$. The solution to (\ref{GHReq}) hence 
reads
\be
\dot{\delta}(t) = \Delta H (t) = \frac{8 \Omega_B(t)}{t} + \frac{C}{t^2}~.
\ee
The homogeneous mode again decays more rapidly than the 
particular solution, so that the dimensionless
quantity $\Delta H / H$ is again asymptotically proportional to $\Omega_B$. 
Instead of eqs.~(\ref{e:RDlimit}),~(\ref{e:RDlimitDe}), we have
\be
\frac{\Delta H}{H} \rightarrow 12 \Omega_B~,\label{e:MDlimit}
\qquad \de(t)-\de(t_{\rm eq}) = \int_{t_{\rm eq}}^t\De H dt \simeq 12\left[
\Omega_B(t_{\rm eq}) -\Omega_B(t_{})\right]\;.
\ee

Let us now take into account a free-streaming component. We want to estimate 
the effect on the photon distribution function
caused by anisotropic expansion in two cases. Case A: 
the neutrinos become non-relativistic \textit{before} photon
decoupling. Case B: the neutrinos become non-relativistic 
\textit{after} photon decoupling. As an approximation,
we assume that this happens instantaneously to all neutrino species, 
such that the contribution of neutrinos to eq.~(\ref{e:lineardelta}) disappears
abruptly. We know that the neutrinos are in fact spread out in momentum space 
and also have a certain spread in the mass spectrum, so in reality this will 
be a  gentle transition. However, we only want to estimate the order of 
magnitude of the effect and are not interested in these details at this point.
More precice numerical results will be presented in sec.~\ref{s:num}.
Let us consider case A first.

\subsubsection{Case A: neutrinos become non-relativistic before photon decoupling}

We know that $\Delta H$ is very nearly zero when the neutrinos become 
non-relativistic. After that, $\Delta H / H$ will start to grow again to
approach the value $6 \Omega_B$ during radiation domination
and $12 \Omega_B$ during matter domination. As boundary condition at photon 
decoupling, we will hence assume
$\Delta H / H = x \Omega_B$ with $x \lesssim 12$.
This number can in principle be computed given the neutrino masses and the 
evolution of the scale factor across matter-radiation equality. We shall
solve the full equations in subsection~\ref{s:num}; here we just want to
understand the results which we obtain there by numerical integration.
The free-streaming component we are interested in now are the photons after 
decoupling. We therefore identify $\delta_\ast = \delta(t_\mathrm{dec})$, 
where $t_\mathrm{dec}$ denotes the instant of photon decoupling.
Furthermore, in eq.~(\ref{e:lineardelta}) we replace
$\Omega_\nu$ by $\Omega_\gamma$, our new free-streaming species. 
With $\Omega_\gamma \propto t^{-2/3}$ in the matter dominated
era, the (not so obvious) analytic solution to eq.~(\ref{e:lineardelta}) is
\be
\label{e:MDsolution}
\delta(t) - \delta(t_\mathrm{dec}) = \frac{15}{4} \frac{\Omega_B}{\Omega_\gamma} + C \lp[f(t) \cos f(t) - \sin f(t)\rp] + D \lp[f(t) \sin f(t) + \cos f(t)\rp]~,
\ee
where we have introduced $f(t) \equiv 4 \sqrt{2 \Omega_\gamma(t)/5}$. 
The time derivative of eq.~(\ref{e:MDsolution}) yields
\be\label{e:deHmga}
\frac{\Delta H}{H} = \frac{16}{5} \Omega_\gamma \lp[C \sin f(t)- D \cos f(t) \rp]~.
\ee
Note that the slowly decaying mode has the same asymptotic behavior as 
(\ref{e:MDlimit}) -- in the matter dominated era, the free-streaming radiation 
can never catch up to the magnetic field, since both fade away too quickly. 
In other words, this means that free-streaming photons are never able to counteract the magnetic field anisotropy in order to isotropize again the Universe, even if they represent the main contribution to the background radiation energy density, and the reason for this is that they decouple only after the end of radiation dominantion.

In order to estimate the value of $\delta$ today
($t_0$), we can simply take the limit of small $\Omega_\gamma(t_0) \ll 1$ 
of (\ref{e:MDsolution}). Correction terms
are suppressed at least by $\sqrt{\Omega_\gamma(t_0)} \sim 10^{-2}$. We find
\be
\delta(t_0) - \delta(t_\mathrm{dec}) \simeq \frac{15}{4} \frac{\Omega_B}{\Omega_\gamma} + D~.
\ee
The constant $D$ is fixed by the boundary conditions at decoupling, given by $\Delta H / H = x \Omega_B$ and
$\delta = \delta(t_\mathrm{dec})$. These boundary conditions translate to
\bea
D &=& \frac{\Omega_B(t_\mathrm{dec})}{\Omega_\gamma(t_\mathrm{dec})} 
\lp[\frac{\sin f(t_\mathrm{dec})}{f(t_\mathrm{dec})} \lp(\frac{5}{16}x - 
\frac{15}{4}\rp) - \frac{5}{16}x \cos f(t_\mathrm{dec}) \rp]\nonumber\\
&=& \frac{\Omega_B(t_\mathrm{dec})}{\Omega_\gamma(t_\mathrm{dec})} 
\lp[-\frac{15}{4}+ \lp(4 + \frac{2 x}{3}\rp) \Omega_\gamma(t_\mathrm{dec}) + 
\mathcal{O} \lp(\Omega^2_\gamma(t_\mathrm{dec})\rp)\rp]~. \label{e:D}
\eea
In order to obtain the essential behavior we have expanded the boundary 
term as a Taylor series in $\Omega_\gamma(t_\mathrm{dec})\ll1$.
Our final result is
\be
\label{e:estimate}
\delta(t_0) - \delta(t_\mathrm{dec}) \simeq \lp(4 + \frac{2 x}{3}\rp) \Omega_B(t_\mathrm{dec})
\lesssim 12 \Omega_B(t_\mathrm{dec}) ~,
\ee
up to corrections suppressed by powers of $\Omega_\gamma(t_\mathrm{dec})$.

In this case, the CMB quadrupole is not affected by the presence of free-streaming 
neutrinos and we obtain the same result as when neglecting their presence,
\be
C_2 \simeq \frac{16\pi}{225}\left[\de(t_0) -\de(t_{\rm dec})\right]^2 
\simeq \frac{768\pi}{75} \Omega_B^2(t_\mathrm{dec}) \simeq 0.1 r^2\,.
\ee

\subsubsection{Case B: neutrinos become non-relativistic after photon decoupling}

In this case, the presence of the neutrino anisotropic stress will
delay the onset of anisotropic expansion until a time $t_m$ when the neutrinos
become effectively non-relativistic. As before, we will ignore that this
is a gradual process and simply assume that one can define some kind of 
``effective'' $t_m$ at which the neutrino anisotropic stress drops to zero.
The full numerical result is given in section~\ref{s:num}.
The effect of anisotropic expansion on the photon distribution function is 
estimated as follows. We assume there is no anisotropic
expansion between photon decoupling and $t_m$. At later times, neutrino 
anisotropic stress can be ignored. The relevant solution
(\ref{e:MDsolution}) is hence obtained with boundary condition $\dot{\delta}(t_m) = 0$. 
Working through the steps
above once again or simply taking the result~(\ref{e:estimate}) with $t_\mathrm{dec} \rightarrow t_m$ and $x \rightarrow 0$, one finds
\be
\label{lightnusol}
\delta(t_0) - \delta(t_\mathrm{dec}) = \delta(t_0) - \delta(t_m) \simeq 4\, \Omega_B(t_m)~.
\ee
Since $\Omega_B$ decays as $a^{-1}$, the effect of anisotropic expansion in 
case B is suppressed by roughly a factor of $a(t_\mathrm{dec}) /(3 a(t_m))$ 
with respect to case A. For light neutrinos with a highest-mass 
eigenstate close to the current lower bound, this factor can be as small as 
$\sim 0.03$, loosening the constraint on a constant magnetic field 
from the CMB temperature anisotropy correspondingly. Constraints coming 
from Faraday rotation are not affected.

Clearly, the heaviest neutrino becomes massive at redshift 
$z_m =m_\nu/T_\nu \gsim 0.04{\rm eV}/T_\nu \simeq  200$.
One might wonder whether isotropization  can be supported even if only one 
neutrino remains massless, since its contribution to the energy density is 
$\Om_{\nu 1} \simeq 0.23\Om_\ga$. The problem is however 
that, as soon as one neutrino species becomes massive, the equilibrium between
the magnetic field and the neutrino anisotropic stresses is destroyed and, as we
have seen under case A, where one still has free streaming photons, it cannot 
be fully re-established in a matter dominated Universe.

\subsection{Numerical solutions}\label{s:num}

In order to go beyond the estimates derived  so far, we have 
solved eqs.~(\ref{EE1}-\ref{EE2}) numerically with cosmological 
parameters corresponding to the current best-fit $\Lambda$CDM model~\cite{LCDM}.
We use cosmological parameters $\Om_\Lambda = 0.73$, $\Om_m = 0.27$ today, where
$\Om_m$ includes a contribution of massive neutrinos\footnote{CMB observations actually
constrain the matter density at decoupling, such that neutrinos with $m_\nu \lesssim 0.3 \mathrm{eV}$,
which are still relativistic at that time, do not contribute to the measurement of $\Om_m$.
However, since their density parameter today is then also very small, their contribution
to the matter density remains practically irrelevant.} which we approximate
by $\Om_\nu h^2 = N_\nu m_\nu / 94 \mathrm{eV}$ with $N_\nu \simeq 3$.
The contribution to the right-hand side of eq.~(\ref{e:deltaH}) from free-streaming neutrinos is 
obtained by integrating eq.~(\ref{e:pressure}) with the full distribution 
function for massive fermions. More precisely, we compute the 
full distribution function to first order in $\delta$ and perform the 
integration numerically, including the neutrino mass as a parameter. We 
begin to integrate deep inside the radiation dominated era, when 
the neutrinos are still relativistic but already free-streaming. The
asymptotic behavior of solution (\ref{e:RDlimitDe})
can be used as initial condition at neutrino decoupling.  
The constraint equation~(\ref{EE1}) provides the remaining initial condition. We then
integrate until the desired time. We define today $t_{0}$ by $a(t_{0})=1$.

\begin{figure*}
\centering   
\includegraphics[width=11cm]{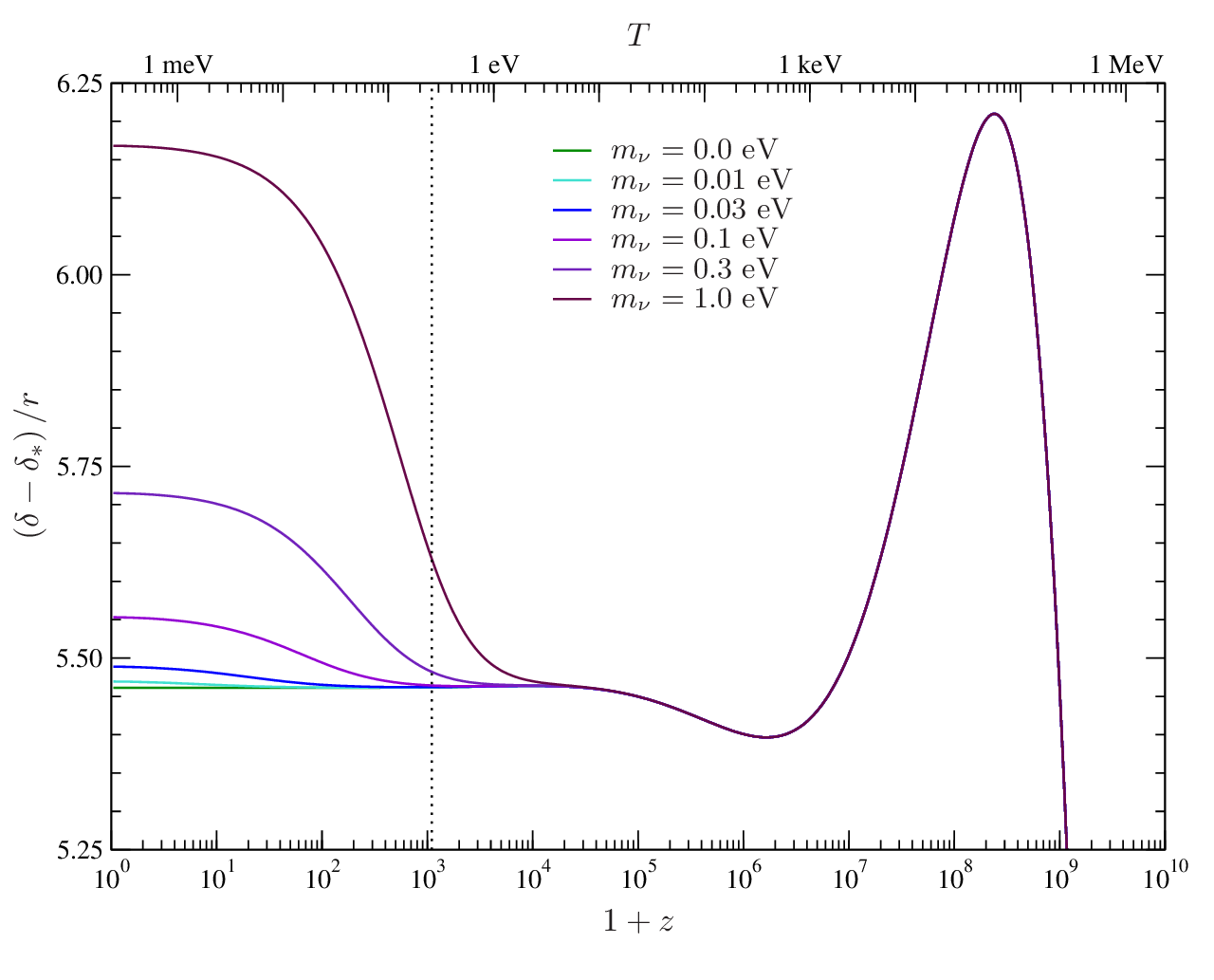}
\includegraphics[width=11cm]{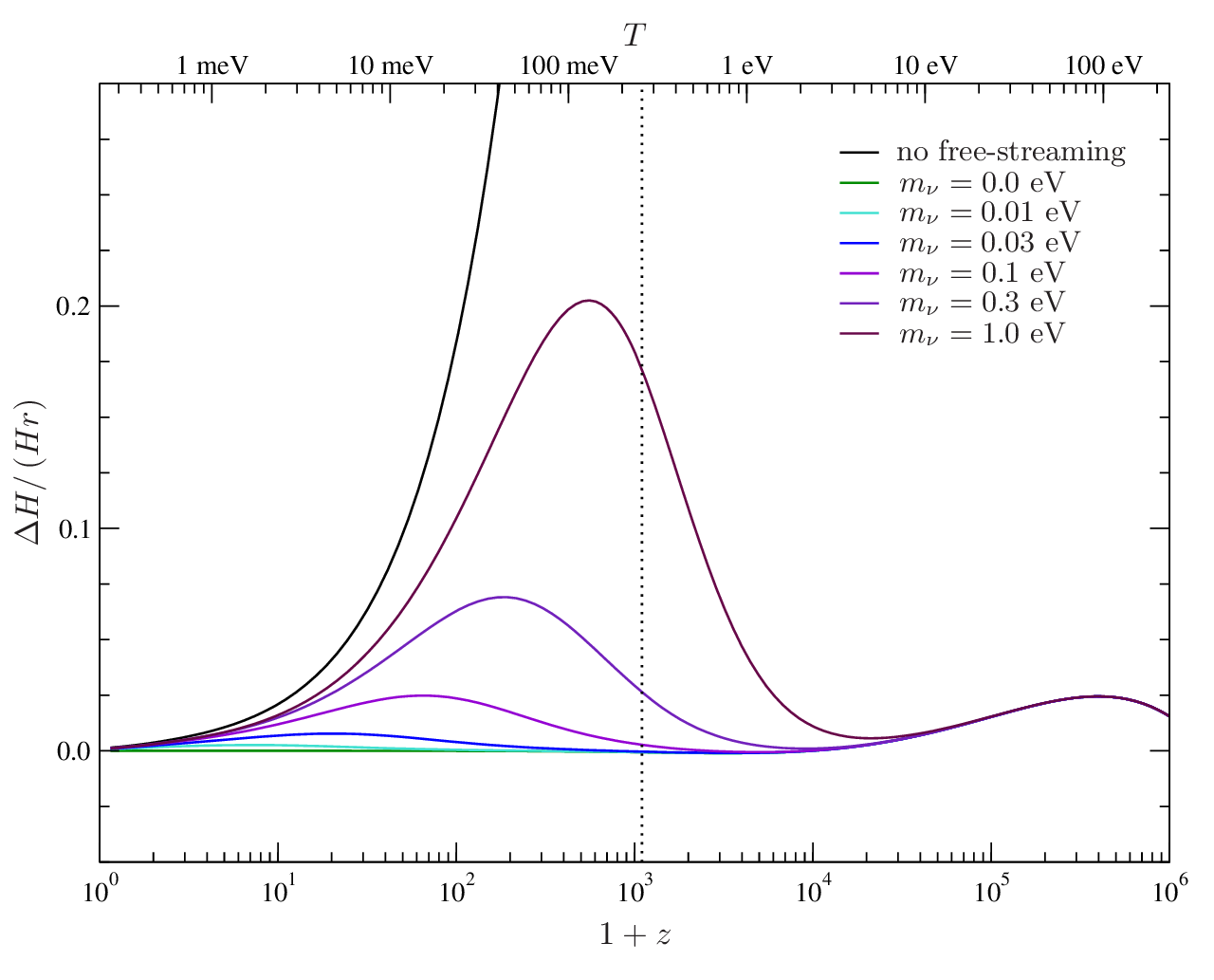}
\caption{Temperature evolution of $\Delta H/H$ and $\de-\delta_\ast$ for different neutrino masses. We chose the initial conditions to be given by $\delta_\ast=0$ at neutrino decoupling. The black solid line in the lower plot represents the temperature evolution of $\Delta H/H$ in the case where only the magnetic field sources the anisotropy and no free-streaming particle is present to compensate this effect. The dotted vertical line indicates the instant of
photon decoupling.
\label{f:nu_final}}
\end{figure*}

In fig.~\ref{f:nu_final}, we present the results of the numerical integration from 
neutrino decoupling until today.  We plot both
$\de-\de_*$ and $\De H/H$ in units of the parameter $r=\Om_B/\Om_\ga$ so that
the plots are valid for arbitrary magnetic field strengths, as long as $r\ll 1$.
After neutrino decoupling, $\delta$ oscillates and reaches its constant value as in eqs.~(\ref{e:RDsolution}),~(\ref{e:MDsolution}), while $\Delta H = \dot{\delta}$ oscillates and decays.
We choose as initial condition $\de=\de_*=0$ at neutrino decoupling.
Once the temperature 
of the Universe reaches the neutrinos mass scale,  neutrino pressure
decreases and they become non-relativistic. At this point, they 
can no longer compensate the anisotropic pressure of the magnetic field, and both 
$\delta$ and $\Delta H$ begin to grow.
However, it is clear from fig.~\ref{f:nu_final} (upper plot) how, once neutrinos become non relativistic after photon decoupling  (case B), the growth of $\de$ is suppressed with respect to case A, where this happens before photon decoupling.
Moreover, the solid black line in the lower plot represents the temperature evolution of $\De H/H$ in the case where only the magnetic field sources the anisotropy: this makes clear how the absence of any free-streaming particle able to counteract the magnetic anisotropic stress leaves  the anisotropy of the Universe free to grow with respect to its value today.

\FIGURE[ht]{ 
\epsfig{width=11cm, file=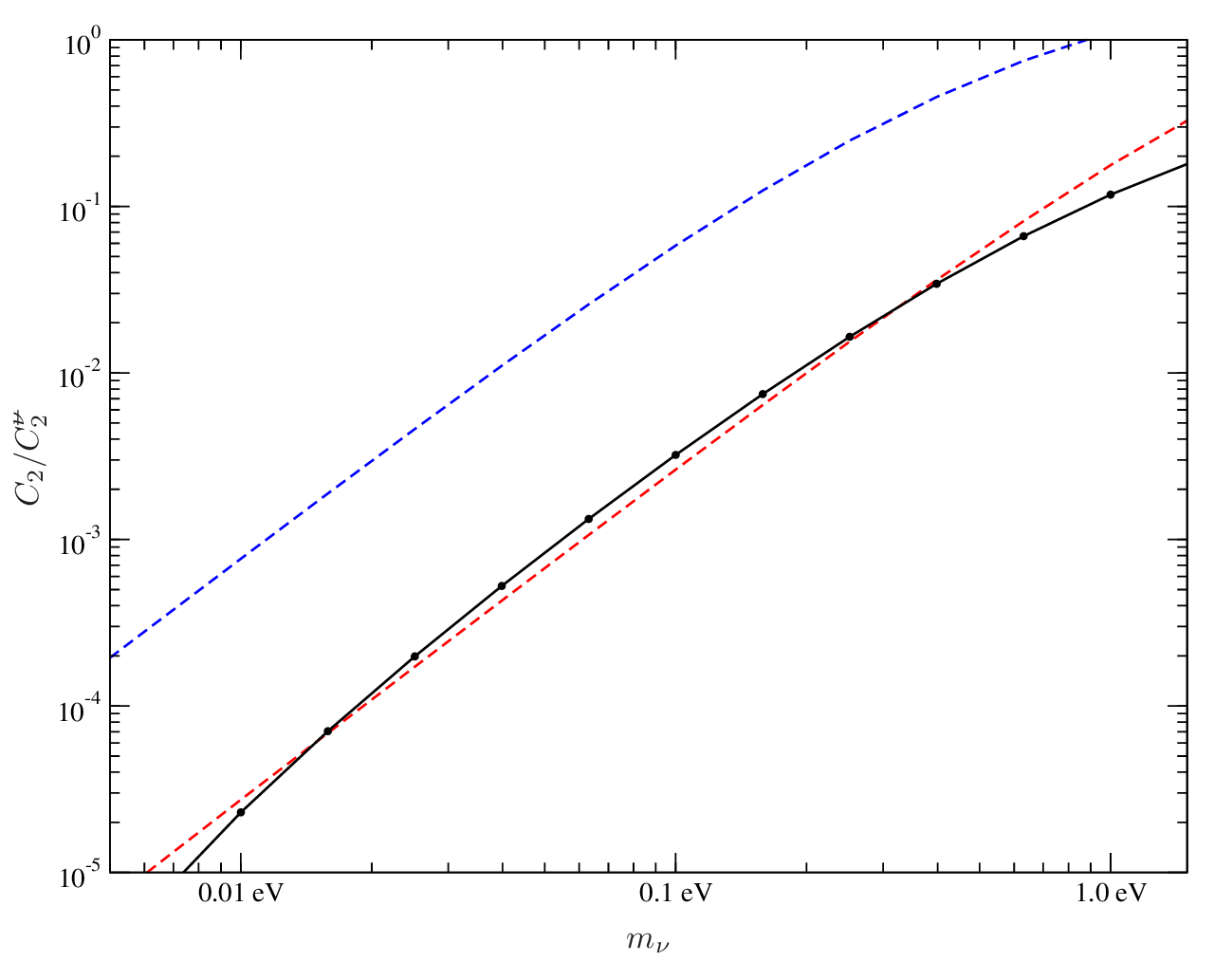}
\caption{Effect of free-streaming neutrinos with different masses on the quadrupole generated
by a homogeneous magnetic field, weighted on the quadrupole obtained without considering the effect of any free-streaming particles.
The solid black line represents the result of the numerical integration, the dashed blue and red lines correspond to our analytical prediction given by 
eq.~(\ref{lightnusol}) for two different choices of $t_m$, the time at which neutrinos are effectively 
non-relativistic (see the text for clarification). }\label{f:quadru}} 

Our quantitative final result is shown  in fig.~\ref{f:quadru}, 
 where we plot the value of the quadrupole generated by a constant 
magnetic field, rescaled by $r^2$, as function of the neutrino mass. 
We weight the final $C_2$ with respect to the quadrupole obtained without considering the isotropization induced by free-streaming particles, in order to underline the relative importance of this effect.
These results 
clearly show that the CMB quadrupole is significantly reduced by neutrino 
free-streaming only if their mass is smaller than the temperature at photon
decoupling, $m_\nu< T_{\rm dec}\simeq0.26\:\mathrm{eV}$. In fact, for neutrino
masses in the range $0.3\:\mathrm{eV} \lesssim m_{\nu} \lesssim 3\:\mathrm{eV}$, 
the quadrupole $C_{2}$ is reduced by less than a factor $100$ from the result
without a free-streaming component, whereas for $0 \lesssim m_{\nu} \lesssim 
0.3\:\mathrm{eV}$, it decreases by several orders of magnitude.
Note, however, that the effect is not negligible even in the former case
with relatively large neutrino masses.
Fig.~\ref{f:quadru} also shows our analytical estimation for the final amplitude of the CMB quadrupole produced by this effect as given by eq.~(\ref{lightnusol}). Of course the value of eq.~(\ref{lightnusol}) depends on the time at which neutrinos become effectively non-relativistic, $t_m$.
Once we choose $t_m$ to be given by the time at which $T=m_\nu$, we overestimate the final quadrupole amplitude still by one order of magnitude (dashed blue line). This is a consequence of the fact that the neutrino distribution function is highly relativistic and therefore it takes a further redshift for them to start  behaving effectively as massive pressureless particles. This has been considered in the more elaborate estimate given by the dashed red line where we fix the time $t_m$ to be given by the time at which $d^3P_\nu/d(\ln T)^3 =0$, i.e.\ the time at which the
pressure reaches the break in the power law. This is in excellent agreement with the numerical results.

\section{A gravitational wave background and other massless free-streaming components in an anisotropic Universe}\label{S:GWs}

From our previous discussion it is evident that any massless free-streaming
particle species $X$ can isotropize the Bianchi~I model with a constant 
magnetic field, if present with sufficient contribution $\Om_X$ already in the 
radiation dominated era. This has to be accounted for if we want to estimate 
the CMB quadrupole 
induced by a homogeneous magnetic field. 

So far we have discussed the standard 
model neutrinos as an example of such a particle. However, also other massless
particles can play this role, for instance gravitons, but also particle 
species outside of the spectrum of the standard model. Interestingly, the 
current bounds on the number of relativistic degrees of freedom during 
nucleosynthesis, often parameterized by the \textit{effective number of additional neutrino species} $\Delta N_\nu$, allow for the possibility that 
such a species could be sufficiently abundant. The present bound on  $N_\nu$ 
from nucleosynthesis is~\cite{numass}
\bea 
N_{\nu} &= & 3.2\pm 1.2~ ,
\nonumber \\
g_* &=& 2+\frac{7N_\nu}{4}\left(\frac{4}{11}
\right)^{4/3}  = 3.36 +(N_\nu-3)\times 0.454 \nonumber \\
&=& 3.36 +(0.2\pm 1.2)\times 0.454 
\quad \mbox{ at } 95\% \mbox{ confidence.} \eea
Here we have taken into account that the photon and neutrino temperatures
are related by $T_\nu = (4/11)^{1/3}T_\ga$~\cite{Ruthbook}.
The effective $g_*$ from $\ga$ and three species of neutrino 
corresponds to $g_*(\ga,3\nu)=3.36$.  This is equivalent to a limit 
on an additional relativistic contribution at nucleosynthesis of 
$\Om_X \lesssim 0.2$. From the solution (\ref{e:RDsolution}) we know that a 
free-streaming relativistic species with a density parameter 
$\Omega_X  \gtrsim 5 / 32 \simeq 0.156$ during the radiation dominated era will 
isotropize expansion within a few Hubble times. Since this species will 
presumably decouple before the neutrinos (otherwise it should have been 
discovered in laboratory experiments), expansion can be isotropic already at 
neutrino decoupling, and thus neither the cosmic neutrino background nor the 
CMB will be affected by anisotropic expansion. In this case therefore, unless we are able 
to detect the background of the species $X$, we will never find a trace of the 
anisotropic stress produced by a homogeneous magnetic field. An interesting 
example are gravitons, which we now want to discuss.

Inflationary models generically predict a background of cosmological 
gravitational waves which are produced from quantum fluctuations during the 
inflationary phase. The amplitude of this background, usually expressed by the 
so-called tensor-to-scalar ratio, $r_T$, has not yet been measured, but for a 
certain  class of inflationary models, forthcoming experiments such as 
Planck might be able to detect these gravitational waves. This is in contrast 
to the cosmic neutrino background, for which there is no hope of direct detection 
with current or foreseeable technology. However, this 
background typically contributes only a very small energy density,
$$ \Om_{\rm GW,\, inf}/\Om_\ga \simeq 10^{-10}r_T ~, \qquad  n_T\lsim 0 \,. $$
Only non-standard inflationary models which allow for $n_T>0$ can contribute a
significant background, see~\cite{blueGWs}.

Gravitational waves can also be produced during phase transitions in the early 
Universe~\cite{ptgw}, after the end of inflation. Such gravitational wave 
backgrounds can easily contribute the required energy density. Let us 
therefore concentrate on this possibility. 

If the highest energy scales of our Universe remain some orders of magnitude 
below the Planck scale, gravitational waves are \textit{never} in thermal 
equilibrium and can be considered as free-streaming radiation throughout the 
\textit{entire history}. Therefore, if the gravitational wave background was
statistically isotropic at some very early time, then any amount of anisotropic
expansion taking place between this initial time and today will affect the 
gravitons in a similar fashion as any other free-streaming component,
and therefore our present gravitational wave background would be anisotropic. 
Loosely speaking, the intensity of gravitational waves would be larger in 
those directions which have experienced less expansion in total since the 
initial time when the gravitational wave background was isotropic.

As we have specified above, with the current limits on $\Delta N_\nu$, the 
density parameter of gravitons $\Omega_\mathrm{GW}$ during nucleosynthesis can 
be as large as $\sim 0.2$. At higher temperatures (that is, at earlier times), 
the number of relativistic degrees of freedom increases (more particle species 
are effectively massless), such that $\Omega_\mathrm{GW}$ at earlier time can 
even be larger\footnote{During a transition from $g_{1}$ 
relativistic degrees of freedom to $g_{2} <g_{1}$, the 
temperature changes from $T_{1}$ to $T_{2}$. Since entropy is conserved 
during the transition we have $g_{1}T_{1}^3=g_{2}T_{2}^3$. Hence 
$\rho_2 =g_{2}T_{2}^4= g_{2}\left[\left(\frac{g_{1}}{g_{2}}\right)^{1/3}
T_{1}\right]^4 =\left(\frac{g_{1}}{g_{2}}\right)^{1/3}\rho_1 >\rho_1$. In 
other words, the energy density
of all species which are still in thermal equilibrium
increases if one reduces the number of 
degrees of freedom at constant entropy.}. 
It is therefore conceivable that gravitons
acquire sufficient anisotropic stress to compensate the magnetic field and 
hence take over the role which neutrinos have played in section \ref{S:nus}. 
As already pointed out, in this case, neither neutrinos nor photons will ever
experience any significant anisotropic expansion, since the Universe remains 
in a Friedmann phase after the gravitons have adjusted to the magnetic field. 
Of course, gravitons remain relativistic for all times and the
mass effect which we discussed for the neutrinos does not occur.

In order to rule out this scenario, it would be very interesting not only to measure 
the background of cosmological gravitational waves but also to determine 
whether or not it shows a quadrupole anisotropy compatible with such a 
compensating anisotropic stress. Or in other words: just as the smallness 
of the CMB quadrupole is a direct indication for isotropic expansion between 
decoupling of photons and today, the smallness of the quadrupole of a 
gravitational wave background would inform us about the isotropy of 
expansion between today and a much 
earlier epoch where this background was generated.

\section{Conclusions}\label{S:con}
In this paper we have studied a magnetic field coherent over very large scales
so that it can be considered homogeneous. We have shown that in the radiation 
dominated era the well known Bianchi~I solution for this geometry is 
isotropized if a free streaming relativistic component is present and 
contributes sufficiently to the energy density, $\Om_X\gtrsim 5/32$. This
is in tune with the numerical finding~\cite{Finelli,ShawLewis,Kerstin} that
the neutrino anisotropic stresses `compensate' large scale magnetic field 
stresses. A perturbative explanation of this effect is attempted in~\cite{CC}.
Here we explain the effect for the simple case of a homogeneous magnetic field:
free streaming of relativistic particles leads to larger redshift, hence smaller 
pressure in the directions orthogonal to the field lines where the magnetic field pressure is 
positive and to smaller redshift, hence larger pressure in the direction 
parallel to the magnetic field, where the magnetic field pressure is
negative. To first order in the difference of the scale factors 
this effect leads to a build up of anisotropic stress in the free streaming
component until it exactly cancels the magnetic field  anisotropic stress.
This is possible since both these anisotropic stresses scale like $a^{-4}$.

In standard cosmology this free-streaming component is given by
neutrinos. However, as soon as neutrinos become massive, their pressure,
$P_\nu \propto a^{-5}$, decays much faster than their energy density, 
$\rho_\nu\propto a^{-3}$, and
the effect of compensation is lost. If this happens significantly
after decoupling, there is still a partial cancellation, but if it happens 
before decoupling, the neutrinos no longer compensate the magnetic field  
anisotropic stress. Furthermore, a component which starts to free-stream 
only in the matter era (like e.g. the photons) does not significantly
reduce the anisotropic stress. Actually, inserting the dominant part of 
the constant $D$ from eq.~(\ref{e:D}) in (\ref{e:deHmga}) one finds
\be \frac{\De H}{H} = 12\Om_B\,, \ee 
like without a free-streaming component.

This cancellation of anisotropic stresses does not affect 
Faraday rotation. A constant magnetic field with amplitude 
$B_0\gsim 10^{-9}$Gauss can therefore be 
discovered either by the Faraday rotation it induces in the CMB~\cite{fara},
or, if a sufficiently intense gravitational wave background exists, by the 
quadrupole (anisotropic stress) it generates in  it.

Finally, Planck and certainly future large scale structure surveys like Euclid will
most probably determine the absolute neutrino mass scale. Once this is known,
we can infer exactly by how much the CMB quadrupole from a constant magnetic 
field is reduced by their presence.

\section*{Acknowledgments} 
We thank Camille Bonvin and Chiara Caprini for discussions. JA wants to thank Geneva 
University for hospitality and the German Research Foundation (DFG) for financial support.
This work is supported by the Swiss National Science Foundation.

\end{document}